\documentclass[traditabstract]{aa}
\usepackage{graphicx}
\usepackage{txfonts}

\usepackage[authoryear]{natbib}
\bibpunct{(}{)}{;}{a}{}{,} 

\newcommand{\Msun}{\ensuremath{{M}_\odot}}
\newcommand{\Rsun}{\ensuremath{{R}_\odot}}
\newcommand{\Mmed}{\ensuremath{{M}_\mathrm{med}}}
\newcommand{\DlogM}{\ensuremath{{\sigma_{M}}}}
\newcommand{\fnemp}{\ensuremath{f_\mathrm{NEMP}}}
\newcommand{\fcnemp}{\ensuremath{f_\mathrm{CNEMP}}}

\begin{document}

\authorrunning{O.R. Pols et al.}

\titlerunning{Nitrogen-enhanced metal-poor stars}

\title{The occurrence of nitrogen-enhanced metal-poor stars:
  implications for the initial mass function in the early Galactic
  halo}

\author{Onno R. Pols\inst{1,2} \and Robert G. Izzard\inst{3,4} \and
  Richard J. Stancliffe\inst{5,6} \and Evert Glebbeek\inst{1,7}}

\institute{
  Department of Astrophysics/IMAPP, Radboud University Nijmegen,
  P.O. Box 9010, NL-6500 GL Nijmegen, The Netherlands. 
  \email{o.pols@astro.ru.nl}
  \and
  Sterrenkundig Instituut Utrecht, P.O. Box 80000,
  NL-3508 TA Utrecht, The Netherlands. 
  \and
  Argelander-Institut f\"ur Astronomie, Auf dem H\"ugel 71,
  D-53121 Bonn, Germany.
  \and
  Institut d'Astronomie et d'Astrophysique, Universit{\'e} Libre de
  Bruxelles, Boulevard du Triomphe, B-1050 Brussels, Belgium.
  \and
  School of Mathematical Sciences, PO Box 28M, Monash University,
  Victoria 3800, Australia.
  \and
  Research School of Astronomy \& Astrophysics, Mount Stromlo Observatory, 
  Cotter Road, Weston Creek ACT 2611, Australia. 
  \and
  Department of Physics and Astronomy, McMaster University, Hamilton,
  Ontario, L8S 4M1, Canada.
}

\date{Received 14 May 2012 / Accepted 19 September 2012}

\abstract{Most carbon-enhanced metal-poor (CEMP) stars are thought to
  result from past mass transfer of He-burning material from an
  asymptotic giant branch (AGB) star to a low-mass companion star,
  which we now observe as a CEMP star. Because AGB stars of
  intermediate mass efficiently cycle carbon into nitrogen in their
  envelopes, the same evolution scenario predicts the existence of a
  population of nitrogen-enhanced metal-poor (NEMP) stars, with
  $\mathrm{[N/Fe]} > 1$ and $\mathrm{[N/C]} > 0.5$. Such NEMP stars
  are rare, although their occurrence depends on metallicity: they
  appear to be more common at $\mathrm{[Fe/H]} < -2.8$ by about a
  factor of 10 compared to less metal-poor stars. We analyse the
  observed sample of metal-poor stars with measurements of both carbon
  and nitrogen to derive firm constraints on the occurrence of NEMP
  stars as a function of metallicity. We compare these constraints to
  binary population synthesis calculations in which we vary the
  initial distributions of mass, mass ratio and binary orbital
  periods. We show that the observed paucity of NEMP stars at
  $\mathrm{[Fe/H]} > -2.8$ does not allow for large modifications in
  the initial mass function, as have been suggested in the literature
  to account for the high frequency of CEMP stars. The situation at
  lower metallicity is less clear, and we do not currently have
  stellar models to perform this comparison for $\mathrm{[Fe/H]} <
  -2.8$. However, unless intermediate-mass AGB stars behave very
  differently at such low metallicity, the observed NEMP frequency at
  $\mathrm{[Fe/H]} < -2.8$ appears incompatible with the top-heavy
  forms of the initial mass function suggested in the literature.}


\maketitle

\section{Introduction}
\label{sec:Introduction}

Carbon-enhanced metal-poor (CEMP) stars make up a large proportion of
the most metal-poor stars in the Galactic halo. Estimates of the
fraction of very metal-poor stars (with $\mathrm{[Fe/H]} < -2$) having
carbon enhancements $\mathrm{[C/Fe]} > 1.0$ range between 10\% and
25\% \citep{2006_Frebel,2006_Lucatello,2012_Carollo}.
Based on their abundance patterns different subgroups are identified
among the CEMP stars \citep{2005_Beers+Christlieb,2010_Masseron}. By
far the most numerous group are also enriched in $s$-process elements,
which make up about 80\% of CEMP stars \citep{2007_Aoki}. We refer
to this group as CEMP-s stars, noting that a certain fraction of these
are also enriched in $r$-process elements (these have been designated
as CEMP-rs stars in the literature). Radial velocity monitoring
suggests that probably all CEMP-s stars are binaries
\citep{2005_Lucatello_binfrac}.  A widely accepted formation scenario
for this group involves pollution by mass transfer in a binary system
from a more massive asymptotic giant branch (AGB) companion, which has
since become a white dwarf.
Of the CEMP stars without strong $s$-process enhancements, one star is
found to be strongly enriched in $r$-process elements, the remainder
having no or only weak overabundances of heavy elements. The latter
group are known as CEMP-no stars. The origin of these stars is less
clear; there is no strong evidence for a large binary fraction among
the CEMP-no stars. They appear to be more common at the lowest
metallicities.  Although they may have followed a similar evolution
path as the CEMP-s stars, it is also possible that they have an
entirely different origin \citep[e.g.][]{2010_Meynet}.

Within the mass transfer scenario, the large proportion of CEMP-s
stars requires the existence of a sufficient number of binary systems
with primary components that have undergone AGB nucleosynthesis.
Several authors (\citealp{2001_Abia,2005_Lucatello_IMF,2007_Komiya})
have therefore argued that a different initial mass function (IMF),
weighted towards intermediate-mass stars, is needed at low
metallicity.
\citet{2005_Lucatello_IMF} find that the standard IMF for the Galactic
disk \citep{1979_Miller+Scalo} contains too few AGB progenitors in the
mass range 1.2--6\,\Msun, assumed to be responsible for carbon and
$s$-element production, to explain the observed fraction of CEMP-s
stars. They propose an IMF with a mild shift towards intermediate-mass
stars is required. 
\citet{2007_Komiya} present a semi-analytical model for binary
population synthesis, based on AGB models in which low-mass stars
(0.8--1.5\,\Msun) at $\mathrm{[Fe/H]}<-2.5$ produce carbon and
$s$-elements by dual shell flashes (see Sect.~\ref{sec:Discussion} for
details). While this increases the parameter space for making CEMP-s
stars considerably, it is barely sufficient to produce a CEMP-s
fraction of more than 10\% with a Galactic-disk IMF. They further
propose that the CEMP-no stars are produced in binary systems with AGB
primaries of 3.5--6\,\Msun. From an assumed CEMP-no/CEMP-s ratio of at
least 1/3, consistent with observations available to them, they find
consistency with both the CEMP-s fraction (10--25\%) and the
CEMP-no/CEMP-s ratio for a top-heavy IMF, with a median mass of
10\,\Msun.
Such proposed changes to the IMF would have profound consequences,
since the IMF affects such diverse quantities as the mass-to-light
ratio of galaxies, the overall production of heavy elements and
galactic chemical evolution.

However, the model calculations on which these conclusions are based
still contain many uncertainties regarding the evolution and
nucleosynthesis of low-metallicity AGB stars, the efficiency of mass
transfer, and the evolution of the surface abundances of the CEMP
stars themselves.
A different approach to the problem of ubiquity of CEMP stars was
taken by \citet{2009_Izzard}. They performed detailed binary
population synthesis calculations to explore whether the observed
CEMP-s fraction at $\mathrm{[Fe/H]} \approx -2.3$ can be explained
within the boundaries of the model uncertainties, while assuming a
standard solar-neighbourhood IMF. Their results confirmed that with
current detailed AGB models, in which dredge-up of carbon occurs only
for masses larger than 1.2\,\Msun, the large CEMP fraction cannot be
explained. However, they also found that efficient dredge-up in
lower-mass AGB stars (0.8--1.2\,\Msun) can substantially increase the
expected CEMP fraction to within the observed range. Therefore the
question of whether CEMP stars provide evidence for a modified IMF at
low metallicity remains open.

Apart from carbon, substantial enhancements of nitrogen with respect
to iron are common among CEMP stars, typically with $\mathrm{[N/C]} <
0$. Detailed AGB nucleosynthesis models of low initial mass ($<
3$--$5\,\Msun$, depending on metallicity) make carbon through the
dredge-up of helium-burning products, but produce very little
nitrogen. On the other hand, AGB stars of higher mass are predicted to
convert the dredged-up carbon into nitrogen by CN cycling at the
bottom of the convective envelope (hot bottom burning, HBB). The
surface abundances of these more massive AGB stars approach the CN
equilibrium ratio of $\mathrm{[N/C]} \approx 2$. Detailed evolution
models of AGB stars \citep{2007_Karakas_yields} indicate that, while
at solar metallicity HBB sets in at $\ga 5\,\Msun$, significantly
lower masses are needed at low metallicity. At $\mathrm{[Fe/H]}=-2.3$
the lower mass limit is found to be $\approx 3\,\Msun$
\citep{2010_Karakas_yields}. By the same mass-transfer scenario that
forms CEMP-s stars, one may thus expect a population of so-called
nitrogen-enhanced metal-poor (NEMP) stars, with enhanced nitrogen
abundances and $\mathrm{[N/C]} > 0.5$.
Note that we thus expect NEMP stars to result from a similar AGB mass
range that \citet{2007_Komiya} suggested to produce CEMP-no stars
(based on models in which no HBB occurs).
Although a few examples of NEMP stars are known, mostly at
$\mathrm{[Fe/H]}<-2.9$, they appear to be very rare
\citep{2007_Johnson}.

In this paper we show that the number of NEMP stars sets an additional
constraint on possible changes to the IMF at low metallicity. In
Sect.~2 we discuss the observational evidence regarding NEMP stars.
Our population synthesis models are described in Sect.~3 and the
results are presented in Sect.~4. We discuss our results in Sect.~5
and in Sect.~6 we give our conclusions.

\section{Nitrogen-enhanced metal-poor stars}
\label{sec:NEMPs}

\citet{2007_Johnson} used the same theoretical considerations as we
gave above to argue for the expected existence of a substantial
population of NEMP stars, which they defined as stars with
$\mathrm{[N/Fe]} > 0.5$ and $\mathrm{[N/C]} > 0.5$. They set out to
search for NEMP stars by determining nitrogen and carbon abundances
for a sample of 21 mildly carbon-enhanced stars. They found no NEMP
stars but confirmed the known trend that $\mathrm{[N/C]}$ values of
CEMP stars tend to be intermediate between what is expected from AGB
stars with masses above and below the hot bottom burning limit. While
their study showed that NEMP stars are rare, and their sample has the
advantage of being homogeneous, it is too small to derive firm
constraints on the frequency of NEMP stars. We therefore performed a
literature search to obtain a more complete overview of the situation.

\begin{figure}[t]
\includegraphics[width=\columnwidth]{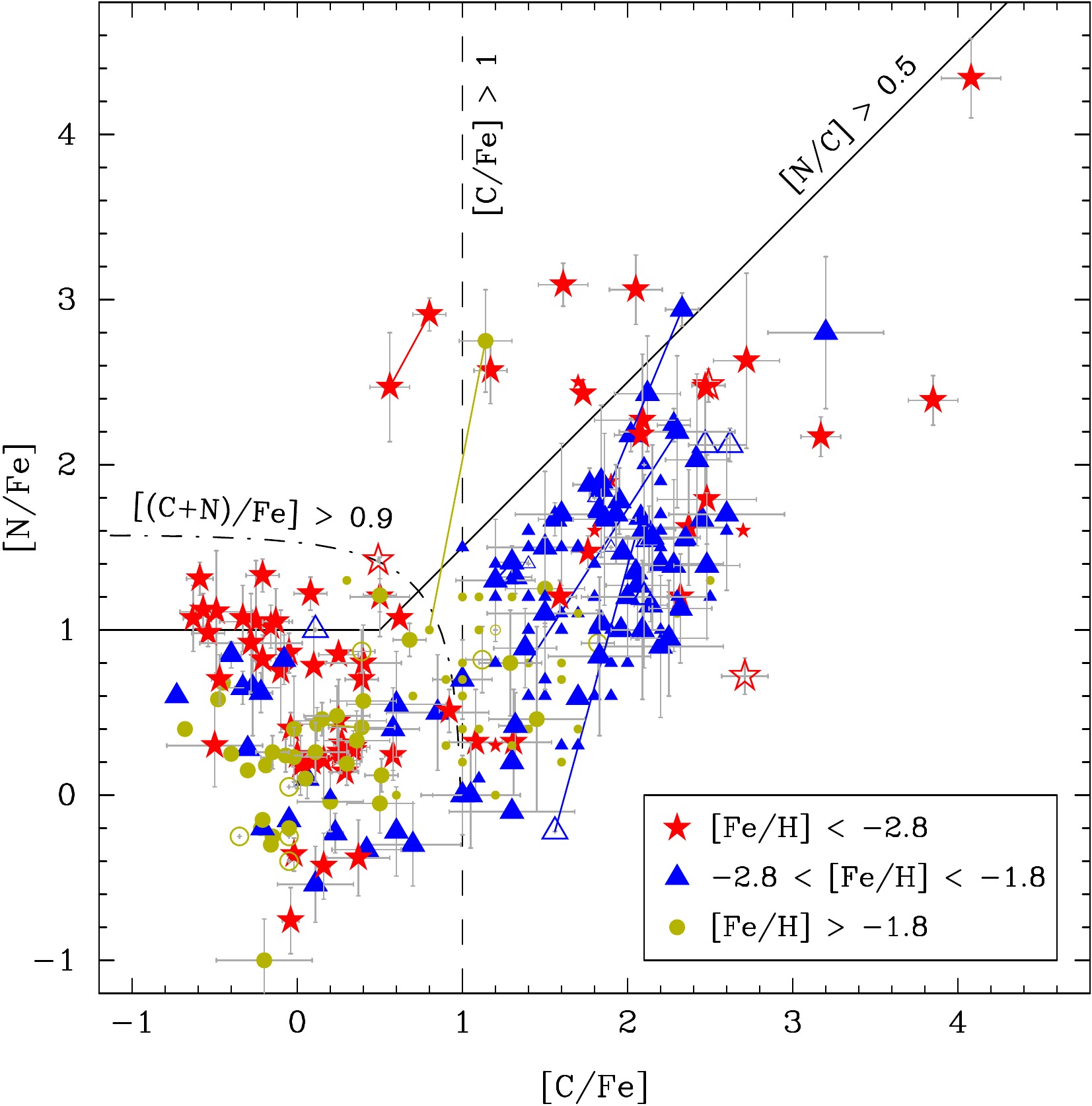}
\caption[]{ Distribution of [N/Fe] versus [C/Fe] for metal-poor stars
  with observed carbon and nitrogen abundances. Star-shaped (red)
  symbols are stars with $\mathrm{[Fe/H]} < -2.8$, triangles (dark
  blue) have $-2.8 \leq \mathrm{[Fe/H]} \leq -1.8$ and circles
  (yellow) have $\mathrm{[Fe/H]} > -1.8$. Giants (stars with $\log g
  \leq 4$) are shown as filled symbols while higher-gravity stars have
  open symbols. Large symbols represent data from
  \citet{2010_Masseron} and the SAGA database; data from
  \citet{2006_Lucatello} is shown with smaller symbols. Multiple data
  for the same star are connected with solid line segments. The dashed
  and solid lines indicate the criteria adopted for CEMP stars and
  NEMP stars, and the dash-dotted line is the alternative criterion
  $\mathrm{[(C+N)/Fe]} > 0.9$ (see text). }
\label{fig:C-vs-N-obs}
\end{figure}

Characterising a star as NEMP requires measurements of both the C and
N abundances, which are only available for a limited number of halo
stars.
In Fig.~\ref{fig:C-vs-N-obs} we show the distribution of [N/Fe] versus
[C/Fe] of 259 metal-poor stars with measurements of both carbon and
nitrogen. Data for 176 stars have been collected in the SAGA database
\citep{2008_Suda_SAGA,2011_Suda} and by \citet{2010_Masseron}.  These
datasets largely overlap but \citeauthor{2010_Masseron} list a few
stars not included in SAGA. Many stars have multiple measurements of
both C and N that are broadly consistent within the observational
errors, in which case we generally take the data from the most recent
publication as listed by \citet{2010_Masseron}. For a few stars with
clearly discrepant measurements, we show two data points connected by
a line segment. Added to this selection are 87 stars, mostly CEMP,
with C and N abundances determined by \citet{2006_Lucatello}, which
were not included in the SAGA database.  We note that the
\citet{2007_Johnson} data are included in SAGA.

\begin{table*}[t]
  \caption[]{%
    Nitrogen-enhanced metal-poor stars, and other possible remnants of
    intermediate-mass binaries (last two entries). }
  \label{tab:NEMPs}
\centering
\begin{tabular}{llrllcllll}
\hline \noalign{\smallskip}
star & [Fe/H] & $\log g$ & [C/Fe] & [N/Fe] & $^{12}$C/$^{13}$C & [Na/Fe] & [Mg/Fe] & [Ba/Fe] & reference \\
\hline \noalign{\smallskip}
\object{CS22949-037}  & $-3.86$ &  1.7 & +1.07 & +2.11 &     & +1.45 & +1.50 &$-0.59$& C08 \\
                      & $-3.97$ &  1.5 & +1.17 & +2.57 &   3 & +2.09 & +1.58 &$-0.58$& D02 \\
\object{CS22960-053}  & $-3.14$ &  2.1 & +2.05 & +3.06 &     &       & +0.65 & +0.86 & A07 \\
                      & $-3.08$ &  2.4 & +1.15 & +1.15 &     &       &       &       & J07 \\
\object{CS29528-041}  & $-3.30$ &  4.0 & +1.61 & +3.09 &     & +1.20 & +0.45 & +0.97 & S06 \\
\object{CS30314-067}  & $-2.85$ &  0.7 & +0.50 & +1.20 &   5 &$-0.08$& +0.42 &$-0.57$& A02 \\
                      & $-2.76$ &  1.0 & +0.25 & +0.50 &     &       &       &       & J07 \\
\object{CS30322-023}  & $-3.39$ &$-0.3$& +0.80 & +2.91 &   4 & +1.29 & +0.80 & +0.52 & M06, M10 \\
                      & $-3.25$ &  1.0 & +0.56 & +2.47 &     & +1.04 & +0.54 & +0.59 & A07 \\
\object{HD25329}      & $-1.8$  &  4.7 & +0.11 & +1.00 &$>$40& +0.24 & +0.59 & +0.11 & G00, G03, F00 \\
\object{HD206983}     & $-0.99$ &  0.6 & +0.50 & +1.21 &   5 &       & +0.26 & +0.92 & M10 \\
\object{HE0400-2030}  & $-1.73$ &  3.5 & +1.14 & +2.75 &     & +0.71 & +0.62 & +1.64 & A07 \\
                      & $-1.7$  &  2.5 & +0.8  & +1.0  &     &       &       &       & L06 \\
\object{HE1031-0020}  & $-2.93$ &  2.2 & +1.73 & +2.43 &     &       & +0.53 & +1.28 & C06 \\
\object{HE1337+0012}\tablefootmark{a}
                      & $-3.20$ &  4.4 & +0.49 & +1.42 &     &$-1.13$& +0.41 &$-0.25$& A06 \\
\object{HE1410+0213}  & $-2.52$ &  2.0 & +2.33 & +2.94 &   3 &       & +0.33 & +0.05 & M10 \\
                      & $-2.23$ &  3.5 & +1.83 & +1.73 &     &       & +0.21 & +0.14 & C06 \\
\object{HE1413-1954}  & $-2.9$  &  4.0 & +1.7  & +2.5  &     &       &       &       & L06 \\
\object{HE2150-0825}  & $-1.3$  &  3.8 & +0.3  & +1.3  &     &       &       &       & L06 \\
\object{HE2253-4217}  & $-2.4$  &  1.6 & +1.0  & +1.5  &     &       &       &       & L06 \\
\hline \noalign{\smallskip}
\object{CS29528-028}  & $-2.86$ &  4.0 & +2.77 &$<$+3.58 &   & +2.68 & +1.69 & +3.27 & A07 \\
\object{SDSS1707+58}  & $-2.52$ &  4.2 & +2.1  &       &     & +2.71 & +1.13 & +3.40 & A08 \\
\hline
\end{tabular}
\tablebib{A02: \citet{2002_Aoki}, A06: \citet{2006_Aoki}, A07:
\citet{2007_Aoki}, A08: \citet{2008_Aoki}, C06: \citet{2006_Cohen},
C08: \citet{2008_Cohen}, D02: \citet{2002_Depagne}, F00:
\citet{2000_Fulbright}, G00: \citet{2000_Gratton}, G03:
\citet{2003_Gratton}, J07: \citet{2007_Johnson}, L06:
\citet{2006_Lucatello}, M06: \citet{2006_Masseron}, M10:
\citet{2010_Masseron}, S06: \citet{2006_Sivarani}.}
\tablefoottext{a}{Also known as G64--12.}
\end{table*}

Following \citet[hereinafter Paper~I]{2009_Izzard}, we designate as
CEMP stars those with $\mathrm{[C/Fe]} \geq 1.0$ and as NEMP stars
those with
\begin{equation}\label{eq:NEMP-criterion}
  \mathrm{[N/Fe]} \geq 1.0 \qquad \mbox{and} \qquad 
  \mathrm{[N/C]} \geq 0.5.
\end{equation} 
Note that these definitions partly overlap; stars that satisfy both
criteria (i.e.\ stars with $\mathrm{[C/Fe]} \geq 1.0$ and
$\mathrm{[N/C]} \geq 0.5$) are designated as CNEMP stars. The adopted
CEMP and NEMP criteria are shown as dashed and solid lines in
Fig.~\ref{fig:C-vs-N-obs}.

Our NEMP criterion~(\ref{eq:NEMP-criterion}) is somewhat more
restrictive on the nitrogen abundance than the definition of
\citet{2007_Johnson}. Application of this criterion to our selection
of data yields 24 NEMP star candidates (see also Table~2 of
Paper~I\footnote{We found 17 stars in Paper~I. The difference is due
  partly to the addition of new data and partly to our use in Paper~I
  of the average observed abundance, while here we use data from a
  single source.}). Ten of these stars, however, have $\mathrm{[C/Fe]}
\la 0.0$ and $\mathrm{[N/Fe]}$ values between 1.0 and 1.3. All these
stars have $\mathrm{[Fe/H]} < -2.8$ and are on the upper part of the
red giant branch (RGB): they belong to the sample of `mixed stars' of
\citet{2005_Spite,2006_Spite}.  Their nitrogen enhancements are very
likely the result of CN cycling and `extra mixing' on the RGB itself,
and not the result of mass accretion of nitrogen-rich material.
Indeed, their combined C + N abundances are only mildly super-solar,
$\mathrm{[(C+N)/Fe]} \la 0.7$.  Our aim of eliminating such stars from
our selection by the criterion $\mathrm{[N/Fe]} \geq 1.0$ was
obviously not completely successful.

An alternative, and perhaps better, criterion to distinguish stars
that have had significant accretion of He-burning products (either
carbon or nitrogen) can be based on the C + N abundance, because
CN-cycling during HBB conserves the total number of C + N nuclei. If
carbon is enhanced by a factor greater than 10 (corresponding to our
CEMP criterion $\mathrm{[C/Fe]} \geq 1.0$) and subsequently cycled
partly into nitrogen, we expect $\mathrm{[(C+N)/Fe]} \geq 0.9$,
assuming \citet{1996_Grevesse} solar-scaled initial abundances. This
criterion is shown as the dash-dotted line in
Fig.~\ref{fig:C-vs-N-obs}. It effectively eliminates stars that are
N-rich owing to internal mixing processes on the RGB.
In this paper, however, we prefer to use our
criterion~(\ref{eq:NEMP-criterion}) because it is consistent with
Paper~I and our model calculations, but we will subject NEMP stars
with $\mathrm{[(C+N)/Fe]} < 0.9$ to critical examination.

If we eliminate the 10 `mixed' RGB stars we are left with 14 NEMP
candidates, for which we list several key properties in
Table~\ref{tab:NEMPs}. However, not all of these stars necessarily
qualify as `genuine' NEMP stars, i.e., stars that are likely to have
been polluted by nitrogen-rich material, possibly from an
intermediate-mass AGB star. Other clues can be obtained from the
$^{12}$C/$^{13}$C ratio, which is expected to be close to its
CN-equilibrium value of 3--4 for material that has gone through HBB,
and from measurements of Na and Mg.
In an AGB star undergoing HBB, Na is produced by the NeNa chain from
the abundant $^{22}$Ne brought to the surface by third dredge-up,
along with Mg if the $^{22}$Ne($\alpha$,n)$^{25}$Mg source is active.
Therefore [Na/Fe] and [Mg/Fe] values greater than zero are expected in
material lost by an intermediate-mass AGB star \citep[e.g.\
see][]{2012_Lugaro}, although we note that Na production by AGB stars
is uncertain owing to the large uncertainty in the
$^{22}$Ne(p,$\gamma$)$^{23}$Na reaction rate \citep{2007_Izzard}.
Clearly, not all NEMP candidates in Table~\ref{tab:NEMPs} for which
$^{12}$C/$^{13}$C, Na and/or Mg are measured show the expected
behaviour. In particular, HD25329 is an unevolved main-sequence star
with $\mathrm{[N/Fe]} = 1.0$ and $\mathrm{[C/Fe]} = 0.1$, and thus a
modest $\mathrm{[(C+N)/Fe]} = 0.5$. Its $^{12}$C/$^{13}$C ratio is at
least 40 which is inconsistent with accretion of material that has
undergone HBB. Although the origin of its high N abundance is unclear,
we should probably exclude this star from the sample of genuine NEMP
stars.  We comment in more detail on the stars in
Table~\ref{tab:NEMPs} in Appendix~\ref{sec:NEMP-individual}.

\begin{table*}[t]
  \caption[]{%
    Incidence of CEMP, NEMP and CNEMP stars among stars with carbon
    and nitrogen measurements. The columns labelled $f_\mathrm{max}$
    give the 90\%, 99\% and 99.9\% upper limits to \fnemp\ and
    \fcnemp\ for the selections with $\log g \leq 4$.}
  \label{tab:NEMP-fractions}
\centering
\begin{tabular}{lccccccc}
  \hline \noalign{\smallskip}
  selection & CEMP & NEMP & CNEMP & \fnemp & $f_\mathrm{max}$(90\% / 99\% / 99.9\%) & 
  \fcnemp & $f_\mathrm{max}$(90\% / 99\% / 99.9\%) \\
  \hline \noalign{\smallskip}
  all &                                          156 & 14 & 8 & 0.086 & & 0.051 \\
  all, $\log g \leq 4$ &                         142 & 12 & 8 & 0.082 & 0.122 / 0.156 / 0.185 & 0.056 & 0.092 / 0.123 / 0.149 \\
  \hline \noalign{\smallskip}
  $\mathrm{[Fe/H]} < -2.8$ &                     ~26 &  8 & 5 & 0.276 & & 0.192 \\
  $\mathrm{[Fe/H]} < -2.8$, $\log g \leq 4$ &    ~24 &  7 & 5 & 0.269 & 0.411 / 0.514 / 0.588 & 0.208 & 0.352 / 0.459 / 0.539 \\
  \noalign{\smallskip}
  $-2.8 \leq \mathrm{[Fe/H]} \leq -1.8$ &        101 &  3 & 2 & 0.029 & & 0.020 \\
  $-2.8 \leq \mathrm{[Fe/H]} \leq -1.8$,
  $\log g \leq 4$ &                               92 &  2 & 2 & 0.022 & 0.057 / 0.088 / 0.116 & 0.022 & 0.057 / 0.088 / 0.116 \\
  \noalign{\smallskip}
  $\mathrm{[Fe/H]} > -1.8$ &                      29 &  3 & 1 & 0.097 & & 0.034 \\
  $\mathrm{[Fe/H]} > -1.8$, $\log g \leq 4$ &     26 &  3 & 1 & 0.107 & 0.223 / 0.316 / 0.390 & 0.038 & 0.142 / 0.229 / 0.304 \\
  \hline \noalign{\smallskip}
  $\mathrm{[Fe/H]} \geq -2.8$ &                  130 &  6 & 3 & 0.045 & & 0.023 \\
  $\mathrm{[Fe/H]} \geq -2.8$, $\log g \leq 4$ & 118 &  5 & 3 & 0.042 & 0.076 / 0.106 / 0.131 & 0.025 & 0.056 / 0.083 / 0.106 \\
  \hline
\end{tabular}
\end{table*}

Fig.~\ref{fig:C-vs-N-obs} demonstrates the paucity of NEMP stars, in
particular among the more metal-rich part of the population. We focus
in particular on the metallicity range $-2.8 \leq \mathrm{[Fe/H]} \leq
-1.8$ because these stars can be best compared to our models at
$\mathrm{[Fe/H]} = -2.3$.
Within this metallicity range (triangles) there are only three
candidate NEMP stars, one of which (HD25329) we can most likely
exclude. The other two are also doubtful cases: HE1410+0213 has an
alternative measurement that gives $\mathrm{[N/C]}<0$, and HE2253-4217
is exactly on the borderline of our criterion with $\mathrm{[N/C]}=0.5$ 
and $\mathrm{[C/Fe]}=1.0$.  Thus, there may be
\emph{no} NEMP stars in the [Fe/H] range that we can directly compare
with our models at $\mathrm{[Fe/H]} = -2.3$. In contrast, NEMP stars
appear to be much more common among the low-metallicity population
($\mathrm{[Fe/H]} < -2.8$, star symbols). However, we are dealing with
low-numbers statistics so we need to take care in drawing such
conclusions. We therefore perform a more careful analysis of the
statistics of NEMP stars.

\subsection{Statistics of CEMP and NEMP stars}
\label{sec:NEMP-stats}

It is difficult to derive an overall fraction of NEMP stars from the
observed samples. The latest release of the SAGA database of
metal-poor stars \citep[dated Oct.\ 31, 2010;][]{2011_Suda} contains
656 stars with measurements of the carbon abundance, of which 110
(about 17\%) classify as CEMP stars. For 148 stars the nitrogen
abundance has also been determined. Out of these, 68 have
$\mathrm{[C/Fe]} > 1$, amounting to a CEMP fraction of 46\% among
stars with both C and N measurements. Apparently there is a strong
bias towards selecting CEMP stars as targets for nitrogen abundance
determinations. The \citet{2010_Masseron} and \citet{2006_Lucatello}
samples of stars with C and N measurements are even more strongly
biased towards CEMP stars.
One reason for this bias may be that in most studies nitrogen
abundances are determined from the CN band, which requires the
presence of a substantial amount of carbon.

On the other hand, among stars for which both C and N abundances
\emph{are} measured there is no reason to suspect a strong bias
towards certain $\mathrm{[N/C]}$ values.  We can thus expect that the
number ratio of NEMP to CEMP stars among these stars is a fairly
unbiased observable.
We designate as \fnemp\ the fraction of NEMP stars among stars that
are either CEMP or NEMP (i.e.\ $\fnemp =
N_\mathrm{NEMP}/[N_\mathrm{CEMP} + N_\mathrm{NEMP} -
N_\mathrm{CNEMP}]$, which avoids counting CNEMP stars twice) such that
$\fnemp \leq 1$ by definition. The common use of CN as a nitrogen
abundance indicator may bias the observed sample of NEMP stars towards
C-rich stars in such a way that non-C-rich NEMP stars are
underrepresented in the observed sample. We therefore also consider
the fraction of CNEMP stars among CEMP stars, which we denote as
\fcnemp\ and which should be unaffected by such a bias.
Another advantage of using the (C)NEMP/CEMP ratio as a constraint on
our model predictions is that it is independent of various model
uncertainties, most notably the binary fraction.  We therefore attempt
to derive \fnemp\ and \fcnemp\ from the 162 stars that are classified
as either CEMP or NEMP from the combined SAGA-Masseron-Lucatello
sample as shown in Fig.~\ref{fig:C-vs-N-obs}.  From the 14 NEMP stars
and 8 CNEMP stars we have found in this sample (Table~\ref{tab:NEMPs})
we thus deduce a nominal $\fnemp=0.086$ and $\fcnemp=0.051$ in the
entire sample.

However, this number deserves closer scrutiny for two reasons.
Firstly, Fig.~\ref{fig:C-vs-N-obs} and Table~\ref{tab:NEMPs}
demonstrate that most NEMP stars have rather low [Fe/H] and the
NEMP/CEMP ratio appears to depend on metallicity. This dependence
becomes clear when we split up the sample into different [Fe/H]
ranges, see Table~\ref{tab:NEMP-fractions}. Among stars with
$\mathrm{[Fe/H]} < -2.8$ both \fnemp\ and \fcnemp\ are about 10 times
higher than in the range $-2.8 \leq \mathrm{[Fe/H]} \leq -1.8$.

Secondly, the numbers quoted in Table~\ref{tab:NEMP-fractions} include
stars of uncertain NEMP status, so we can at best derive upper limits
to the NEMP/CEMP fraction from the observations. Since we are dealing
with very small numbers, especially when we consider sub-samples in
different metallicity ranges, we cannot simply take the nominal
fractions in Table~\ref{tab:NEMP-fractions}.  Instead we should ask
the question: given a true fraction $f$ of NEMP stars among CEMP
stars, what is the probability of finding at most $N$ NEMP stars in an
observed sample of $N_\mathrm{tot}$ CEMP stars? Assuming a purely
stochastic process, this probability is $P(\leq N) \equiv
\sum_{n=0}^{N} P_\mathrm{b} (n, N_\mathrm{tot}, f)$, where
$P_\mathrm{b} (n, N_\mathrm{tot}, f)$ is the binomial probability of
drawing $n$ NEMP stars out of a sample of $N_\mathrm{tot}$ given a
true fraction $f$. Since $P(\leq N)$ is a decreasing function of $f$,
if we find $P(\leq N)$ to be, say, 10\% for a certain value of $f$, we
can exclude NEMP/CEMP fractions larger than $f$ with 90 per cent
confidence. In this way we can compute upper limits to $f$ for certain
confidence limits, i.e.\ 90\%, 99\% and 99.9\%.  These upper limits
are given in Table~\ref{tab:NEMP-fractions} for the different
subsamples and for both \fnemp\ and \fcnemp.

As in \citet{2009_Izzard} we limit our analysis to stars with $\log g
\leq 4$, i.e.\ subgiants and giants. This facilitates the comparison
with our model calculations and does not introduce significant biases
in the numbers we derive, as can be verified in
Table~\ref{tab:NEMP-fractions}.
We further limit the selection to stars with $-2.8 \leq
\mathrm{[Fe/H]} \leq -1.8$, which is the relevant metallicity range
for comparison to our models at $\mathrm{[Fe/H]} = -2.3$. The SAGA
database contains 287 stars with C measurements in this range, out of
which 59 are CEMP stars. This amounts to a CEMP fraction\footnote{%
  In \citet{2009_Izzard} we found a larger CEMP fraction, 31\%, from a
  similar selection on $\mathrm{[Fe/H]}$ and $\log g$. However, the
  dataset used in that paper combined the SAGA database with a
  selection of the data from \citet{2006_Lucatello} which is strongly
  biased towards CEMP stars.} of 20.6\%.  Adding the additional CEMP
stars from \citet{2010_Masseron} and \citet{2006_Lucatello} yields 92
CEMP stars and 2 NEMP stars, both of which are also (marginally)
CNEMP. Despite a nominal \fnemp\ of 0.022 in this range, we cannot
exclude real fractions up to 0.088 with more than 99\% confidence.

The NEMP/CEMP ratio appears to be larger at both lower and higher
metallicity. The difference at lower metallicity is likely to be real,
since at least 5 out of the 7 NEMP stars at $\mathrm{[Fe/H]} < -2.8$
appear to be genuine. We can compute the probability that we observe
\emph{at least} 5 NEMP stars among 26 CEMP+NEMP stars given a real
fraction $f$ in a similar way, and find we can exclude $f<0.097$ with
90\% confidence and $f<0.052$ with 99\% confidence. On the other hand,
the status of the 3 NEMP stars at $\mathrm{[Fe/H]} > -1.8$ is less
certain and the high nominal NEMP/CEMP ratio could just be due to
chance and small-number statistics. We can take these stars into
account by combining them with the $-2.8 \leq \mathrm{[Fe/H]} \leq
-1.8$ sample, see the lower line of Table~\ref{tab:NEMP-fractions}. We
then find upper limits to \fnemp\ that are only slightly larger than
from the $-2.8 \leq \mathrm{[Fe/H]} \leq 1.8$ sample by itself, while
the limits to \fcnemp\ are even slightly lower.

We conclude by noting that the SAGA database is by no means a
statistically complete sample.  Bright giants are clearly
overrepresented with respect to fainter giants and turnoff stars,
compared to what is expected from the relative lifetimes of these
phases \citep{2009_Izzard}. However, neither the observed sample nor
our model results show a strong dependence of the CEMP fraction on
evolution state (represented e.g.\ by the effective gravity).
Therefore the results we describe below are probably not greatly
affected by this selection effect.

\section{Population synthesis models}
\label{sec:Models}

We have simulated populations of metal-poor halo stars in binary
systems using the rapid synthetic binary nucleosynthesis code of
\citet{2004_Izzard,2006_Izzard}. The algorithm and the assumptions
made in these simulations are discussed extensively in Paper~I. We
adopt a metallicity $Z=10^{-4}$ ($\mathrm{[Fe/H]} = -2.3$) and select
stars with ages between 10 and 13.7\,Gyr (roughly corresponding to the
age of the halo) and $\log g \leq 4.0$.  Our synthetic populations
contain only binaries (see Sect.~\ref{sec:Discussion} for a discussion
of the binary fraction).
However, the predicted NEMP/CEMP fraction is independent of the binary
fraction, assuming both CEMP and NEMP stars are formed through binary
mass transfer.

We consider models A, G and H from 
Paper~I, to which we refer for details. Model A is our default model,
where we parameterize AGB evolution, including the efficiency of third
dredge-up (TDU), according to the detailed models of
\citet{2002_Karakas}.  This yields C-rich AGB stars and thus CEMP
progenitors for initial primary masses $M_1 > 1.2\,\Msun$. Hot bottom
burning produces N-rich AGB stars for $M_1 > 2.7\,\Msun$ in these
models.
In Model G we assume much more efficient TDU in low-mass AGB stars by
reducing the minimum core mass and envelope mass required for
dredge-up, as described in Paper~I.  This results in almost all stars
with initial masses 0.85--2.7\,\Msun\ becoming C-rich AGB stars and
thus potential CEMP progenitors. On the other hand, more massive stars
-- and thus NEMP progenitors -- are not affected by this choice.  We
note that more efficient TDU is also required to match the carbon-star
luminosity functions in the Magellanic Clouds \citep{2004_Izzard}.
Models A and G both assume efficient thermohaline mixing of the
accreted matter throughout the stellar envelope of the secondary.
Model H is similar to Model G but we do not allow thermohaline mixing,
such that accreted material remains at the surface of the secondary
with unchanged abundances until first dredge-up occurs.

In this study, in addition to varying the main assumptions about the
input physics, we also vary the initial binary parameter distributions
as detailed below. We have no direct information about these
distributions in the halo at low metallicity, and must necessarily
make some rather crude assumptions.  Models 1--4 are alternative
representations of the binary population in the solar neighbourhood,
with which we test the sensitivity of our results to uncertainties in
these distributions.

\begin{itemize}

\item[1.]{} Default model, as in Paper~I: the initial primary masses
  $M_1$ are distributed according to the solar neighbourhood IMF as
  derived by \citet{1993_Kroupa}, the initial separations are drawn
  from a flat distribution in $\log a$ (with $a$ between 3 and
  $10^5\,\Rsun$) and the initial mass ratios from a flat distribution
  in $q=M_2/M_1$. The IMF and mass-ratio distribution for this model
  are shown as solid lines in Fig.~\ref{fig:map-q-m1}b and c.  A flat
  distribution in $\log a$ provides the best description of the binary
  population of young OB associations \citep{2007_Kouwenhoven}, and
  thus represents intermediate-mass binary populations in the Galaxy
  (with $M_1 \ga 3\,\Msun$).

\item[2.]{} Primary masses and mass ratios are distributed as in Model
  1, but orbital periods are drawn from the distributions derived by
  \citet[hereafter DM91]{1991_Duquennoy+Mayor} for the local
  population of G dwarfs (with $M_1 \sim 1\,\Msun$), i.e.\ a
  log-normal period distribution with a broad peak at 170 years.

\item[3.]{} As model 2, but here also the mass ratios are drawn from
  the distribution proposed by DM91, i.e.\ a Gaussian mass-ratio
  distribution with a broad peak at $q=0.23$. This is shown as the
  dashed line in Fig.~\ref{fig:map-q-m1}c.

\item[4.]{} Primary masses are taken from an alternative IMF for the
  solar neighbourhood by \citet{1979_Miller+Scalo}, described by a
  log-normal distribution,

  \begin{equation}\label{eq:IMFlognormal}
    \frac{dN}{d\log M_1} \propto \exp 
    \Big[- \frac{( \log M_1 - \log\Mmed )^2}{2\,\DlogM^2} \Big],
  \end{equation}
  with a median mass of $\Mmed=0.1$\,\Msun\ and a width $\DlogM =
  0.67$, while separations and mass ratios are distributed as in
  Model~1.

\end{itemize}

\noindent
Models 5 and 6 test the effect of two alternative forms of the IMF
that have been suggested in the literature.

\begin{itemize}
\item[5.]{} As Model 4, but with a modified log-normal IMF
  (Eq.~\ref{eq:IMFlognormal}) with $\Mmed=0.79$\,\Msun\ and $\DlogM =
  0.5$, as proposed by \citet{2005_Lucatello_IMF}. This IMF is shown
  as a dashed line in Fig.~\ref{fig:map-q-m1}b.
\item[6.]{} As Model 4 using a log-normal IMF with a much larger
  median mass of 10\,\Msun\ and a narrower width, $\DlogM = 0.33$, as
  proposed by \citet{2007_Komiya}.
\end{itemize}

We also test the effect of varying the IMF parameters in
Eq.~(\ref{eq:IMFlognormal}), \Mmed\ and \DlogM, in a continuous manner
between 0.1\,\Msun\ and 20\,\Msun\ and between 0.1 and 1.0,
respectively.

\section{Results}
\label{sec:Results}

\begin{figure}[t]
\includegraphics[width=\columnwidth]{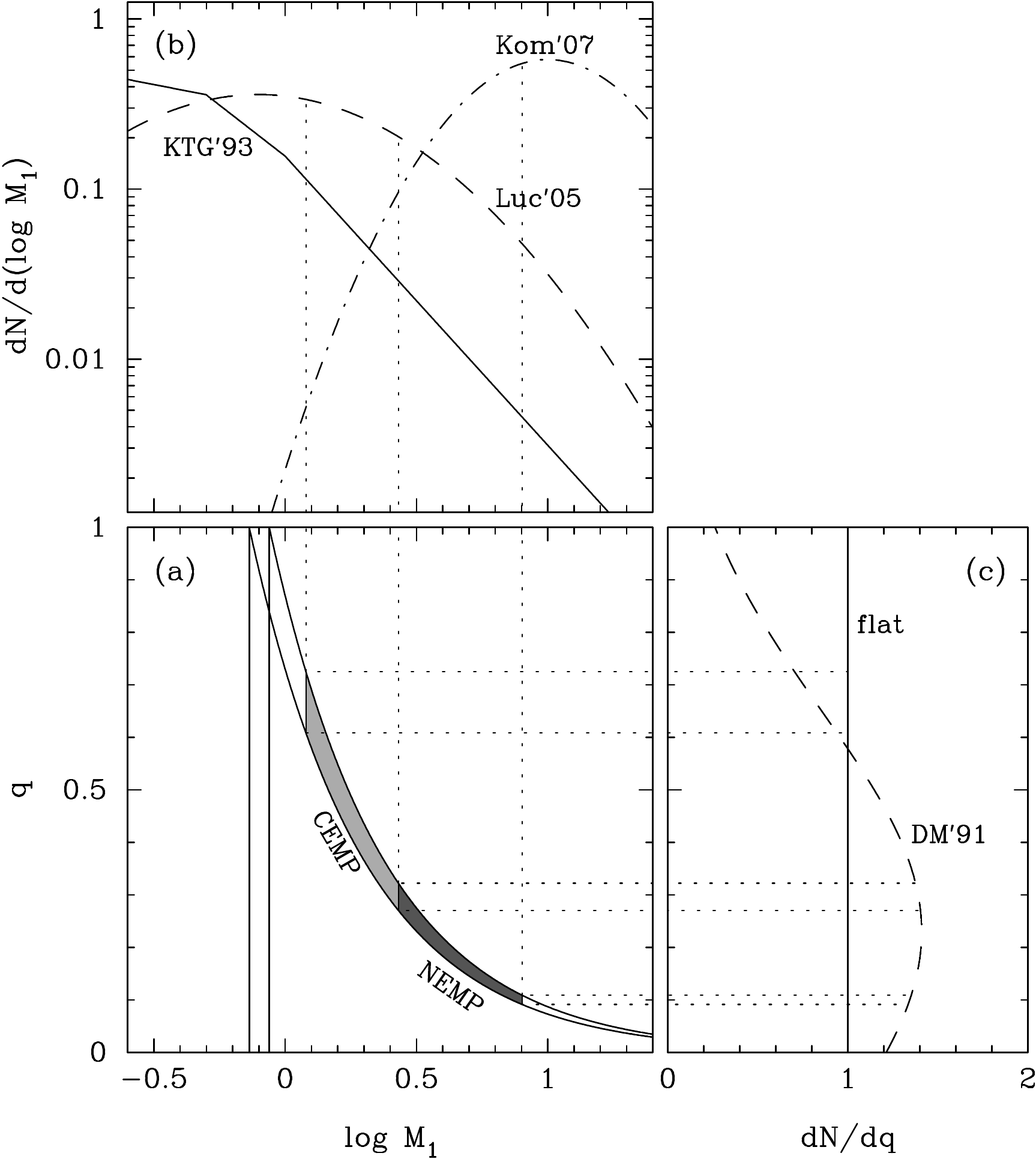}
\caption[]{(a) Schematic regions in the parameter space of initial
  primary mass $M_1$ and mass ratio $q$ for the progenitors of EMP
  stars ($M\approx0.8\,\Msun$, between the solid lines) and of CEMP
  and NEMP stars (light and dark shading, respectively); see text for
  details. (b) Three examples of the initial primary mass function
  \citep{1993_Kroupa,2005_Lucatello_IMF,2007_Komiya}. (c) Comparison
  of two initial mass-ratio distributions. }
\label{fig:map-q-m1}
\end{figure}

Figure~\ref{fig:map-q-m1}a shows in a schematic way where the
progenitors of CEMP and NEMP stars are located in the parameter space
of initial primary mass $M_1$ and initial mass ratio $q$. The
constraints on age and gravity mean that we select stars with masses
in a narrow interval around 0.8\,\Msun. These stars are either primary
components in binaries (indicated by the vertical lines in
Fig.~\ref{fig:map-q-m1}a) or the original secondaries in binaries
where the more massive primary has evolved into a compact remnant. The
latter have $M_2 = q M_1 \approx 0.8\,\Msun$, which defines a small
range of mass ratios $\Delta q \propto 1/M_1$ for each $M_1$ (shown by
the curved lines)\footnote{%
  Note that in Fig.~\ref{fig:map-q-m1} the indicated width $\Delta M$
  of the mass interval around 0.8\,\Msun\ is arbitrary and only for
  illustration purposes; this cancels out when calculating stellar
  number ratios as long as $\Delta M \ll M$. }.  The CEMP and NEMP
progenitors lie in the grey-shaded regions in
Fig.~\ref{fig:map-q-m1}a, where the mass limits shown correspond to
Model~A.  Only binaries with orbital periods in a certain range evolve
into CEMP or NEMP stars; this is not shown, but our models indicate
that this period range is not very sensitive to $M_1$.  Another
simplification in Fig.~\ref{fig:map-q-m1} is that we have ignored
accretion: in reality the initial mass ratios of CEMP and NEMP
progenitors are somewhat smaller because the secondaries accrete up to
0.1\,\Msun\ of primary material.

This figure illustrates several effects that are visible in the
results we obtain. First, both the CEMP fraction among stars with $M
\approx 0.8\,\Msun$ and the NEMP/CEMP ratio depend on the shape of the
IMF, both increasing as the median mass of stars formed increases.
Second, since $\Delta q \propto 1/M_1$, the fraction of binary systems
at a certain $M_1$ that evolve into a CEMP or NEMP star decreases with
primary mass. For a flat mass-ratio distribution this fraction is, to
first order, inversely proportional to $M_1$. This effect results in a
smaller NEMP/CEMP ratio than one would expect on the basis of the IMF
alone.
Third, NEMP progenitor systems have smaller initial mass ratios ($0.1
\la q \la 0.3$) than CEMP progenitors ($0.3 \la q \la 0.7$ for
Model~A). Therefore the NEMP/CEMP ratio is sensitive to the initial
mass-ratio distribution. On the other hand, because the orbital
periods of CEMP and NEMP progenitors are similar, the NEMP/CEMP ratio
is \emph{not} sensitive to the initial period distribution.

\begin{table}[t]
  \caption[]{%
    Predicted number fractions of CEMP, NEMP and CNEMP stars among
    halo stars at $\mathrm{[Fe/H]} \sim -2.3$ and $\log g < 4.0$. The
    last two columns give the fraction \fnemp\ of NEMP stars among
    stars that are either CEMP or NEMP and the fraction \fcnemp\ of
    CNEMP stars among CEMP stars.}
  \label{tab:results}
\centering
\begin{tabular}{crrlcc}
\hline \noalign{\smallskip}
model & CEMP & NEMP & CNEMP & \fnemp & \fcnemp \\ 
\hline \noalign{\smallskip}
A1 & 2.33\,\% & 0.27\,\% & 0.10\,\% & 0.107 & 0.041 \\ 
A2 & 2.75\,\% & 0.34\,\% & 0.11\,\% & 0.114 & 0.041 \\ 
A3 & 3.50\,\% & 0.54\,\% & 0.18\,\% & 0.139 & 0.051 \\ 
A4 & 3.18\,\% & 0.48\,\% & 0.17\,\% & 0.137 & 0.051 \\ 
A5 & 4.87\,\% & 1.04\,\% & 0.36\,\% & 0.188 & 0.074 \\ 
A6 & 13.3\,\% & 19.7\,\% & 5.4\,\%~ & 0.714 & 0.406 \\ 
\hline \noalign{\smallskip}
G1 & ~9.5\,\% & 0.27\,\% & 0.10\,\% & 0.027 & 0.011 \\ 
G2 & 11.1\,\% & 0.34\,\% & 0.12\,\% & 0.030 & 0.011 \\ 
G3 & ~8.5\,\% & 0.53\,\% & 0.19\,\% & 0.060 & 0.023 \\ 
G4 & 10.6\,\% & 0.48\,\% & 0.18\,\% & 0.044 & 0.017 \\ 
G5 & 12.7\,\% & 1.04\,\% & 0.39\,\% & 0.078 & 0.031 \\ 
G6 & 15.6\,\% & 19.7\,\% & 5.8\,\%~ & 0.667 & 0.372 \\ 
\hline \noalign{\smallskip}
H1 & 15.8\,\% & 0.43\,\% & 0.30\,\% & 0.027 & 0.019 \\
H2 & 18.7\,\% & 0.55\,\% & 0.38\,\% & 0.029 & 0.021 \\
H3 & 15.0\,\% & 0.87\,\% & 0.60\,\% & 0.057 & 0.040 \\
H4 & 17.9\,\% & 0.77\,\% & 0.53\,\% & 0.042 & 0.030 \\
H5 & 22.1\,\% & 1.68\,\% & 1.16\,\% & 0.074 & 0.052 \\
H6 & 43.5\,\% & 35.8\,\% & 22.8\,\% & 0.634 & 0.525 \\
\hline
\end{tabular}
\end{table}

\begin{figure}[t]
\includegraphics[width=\columnwidth]{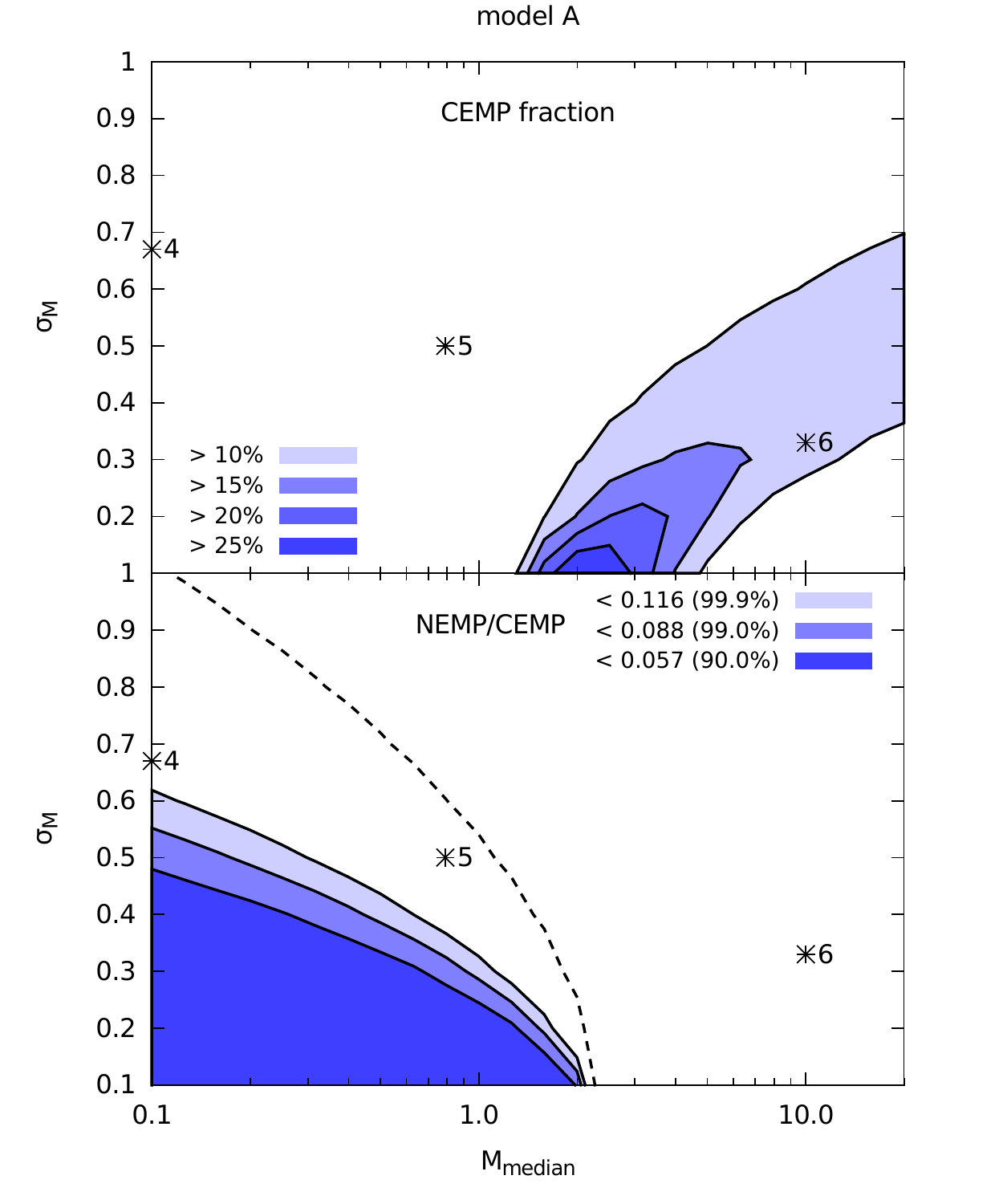}
\caption[]{Dependence of the CEMP fraction (upper panel) and the
  fraction \fnemp\ of NEMP stars among all CEMP+NEMP stars (lower
  panel) for our default model A, as a function of the median mass
  \Mmed\ and dispersion \DlogM\ for a log-normal IMF. Contours in the
  upper panel correspond to CEMP fractions between 10\% and 25\%,
  which spans the range of observed values. The solid contours in the
  lower panel correspond to the 90\%, 99\% and 99.9\% confidence upper
  limits on \fnemp\ in the range $-2.8 \leq \mathrm{[Fe/H]} \leq -1.8$
  and $\log g \leq 4$, while the dashed line shows the 99\% upper
  limit on \fcnemp\ (see Table~\ref{tab:NEMP-fractions}). The star
  symbols labelled 4, 5 and 6 indicate the IMFs of
  \citet{1979_Miller+Scalo}, \citet{2005_Lucatello_IMF} and
  \citet{2007_Komiya}.}
\label{fig:IMF-modelA}
\end{figure}

\begin{figure}[t]
\includegraphics[width=\columnwidth]{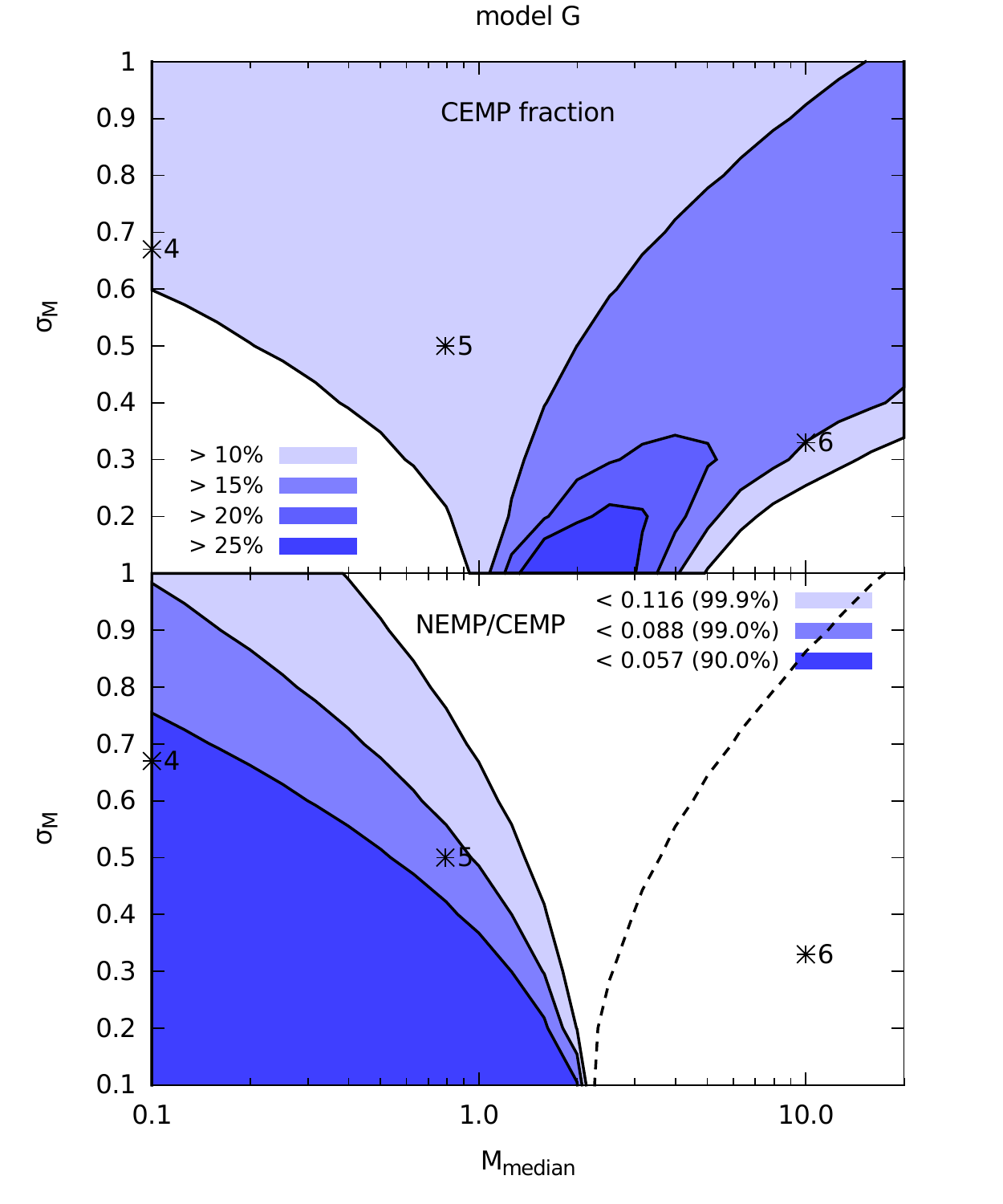}
\caption[]{Same as Fig.~\ref{fig:IMF-modelA} for model G, with
  enhanced dredge-up in low-mass AGB stars.}
\label{fig:IMF-modelG}
\end{figure}

\begin{figure}[t]
\includegraphics[width=\columnwidth]{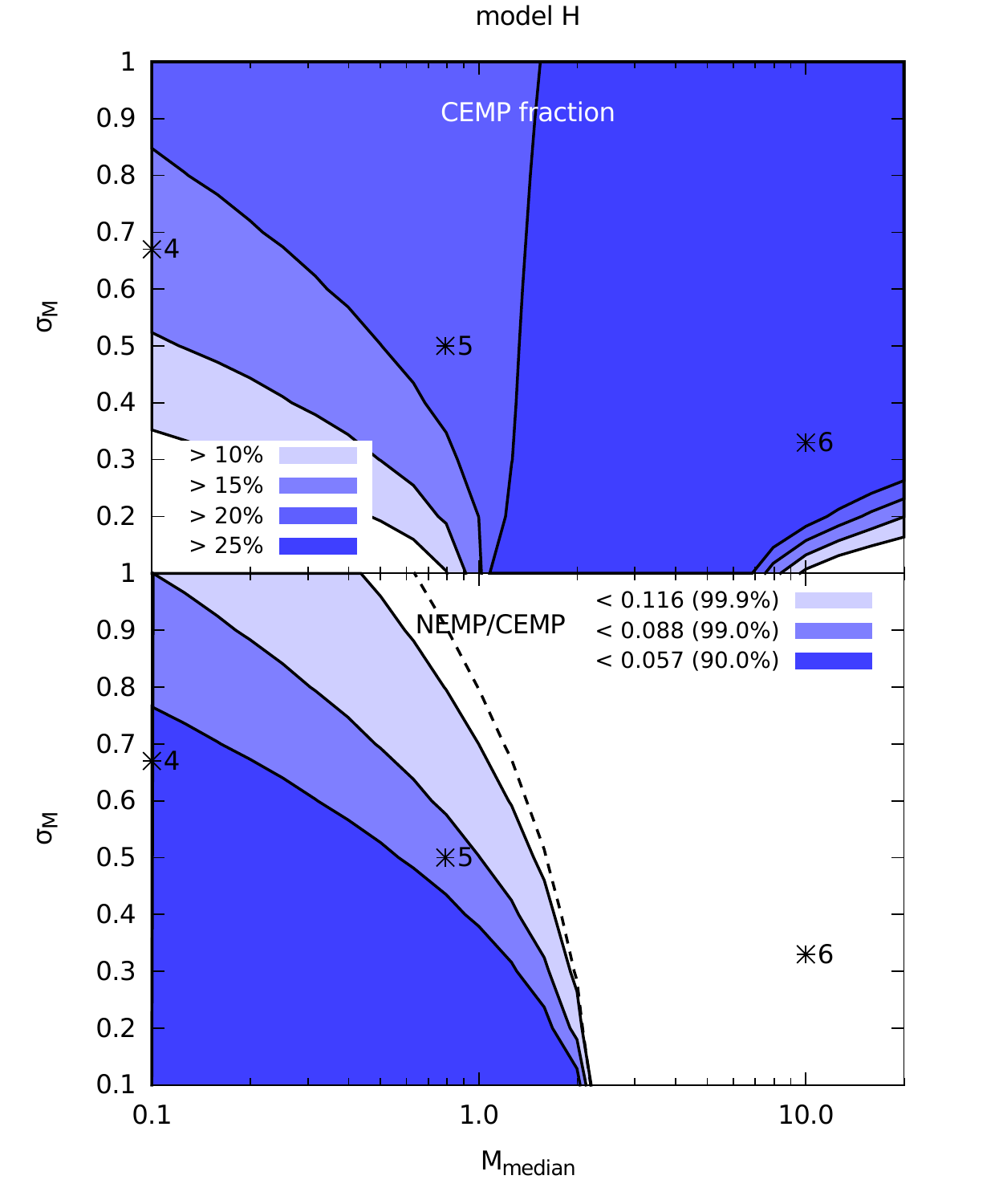}
\caption[]{Same as Fig.~\ref{fig:IMF-modelA} for model H, without
  thermohaline mixing.}
\label{fig:IMF-modelH}
\end{figure}

In Table~\ref{tab:results} we give the number fractions of CEMP, NEMP
and CNEMP stars resulting from our models, for various assumptions
regarding the physical ingredients (models A, G and H) and the initial
distributions of binary parameters (1--6).  We also compute the
(C)NEMP/CEMP fractions \fnemp\ and \fcnemp\ from these models for
comparison with the observational constraints derived in
Sect.~\ref{sec:NEMP-stats}.  Since our models are for $\mathrm{[Fe/H]}
= -2.3$, this comparison should strictly be made only with the
subsample having $-2.8 \leq \mathrm{[Fe/H]} \leq -1.8$.
A substantial fraction of NEMP stars in our models are CNEMP, because
even though intermediate-mass AGB stars convert most dredged-up carbon
into nitrogen by HBB, these stars still produce large amounts of
carbon. This material can be further diluted by thermohaline mixing in
the secondary and therefore in model sets A and G a smaller proportion
of NEMP stars are CNEMP than in model set H, in which no such dilution
occurs until first dredge-up.

The results for models with default binary parameter distributions
(A1, G1 and H1) are discussed extensively in Paper~I.  Our default
model (A1) fails to account for the large observed CEMP frequency,
while Model~G1 yields an increase of the CEMP fraction to almost 10\%,
substantially closer to the observed value.  This corresponds to
decreasing the lower primary mass limit of CEMP progenitors in
Fig.~\ref{fig:map-q-m1} to about 0.85\,\Msun.  Inhibiting thermohaline
mixing increases the CEMP frequency to more than 15\% (Model H1). In
all these default models the small overall NEMP fraction ($<0.5\%$) is
consistent with the observations, and Models~G1 and H1 with enhanced
dredge-up can comfortably reproduce the observed small \fnemp. Based
on the considerations in Sect.~\ref{sec:NEMP-stats}, Model~A1 can be
rejected with more than 99\% confidence on the basis of its \fnemp\
value -- but this is entirely due to the small CEMP fraction in this
model.
If we consider only CNEMP stars, the constraints are weaker and also
Model~A1 (with $\fcnemp=0.041$) cannot be excluded with more than 90\%
confidence.

\subsection{Variations in parameter distributions of the local binary
  population}

Distributions 2 and 3 give some insight into the dependence of our
results on uncertainties in the local (solar-neighbourhood) binary
parameter distributions. Model~A2 shows a modest increase in both the
number of CEMP stars and NEMP stars compared to Model~A1, because the
peak in the \citet{1991_Duquennoy+Mayor} period distribution coincides
with the period range in which mass transfer is most effective.  This
is partly compensated by the wider range of periods in the DM91
distribution. Similarly modest effects are seen when considering
Models~G2 and H2. \fnemp\ and \fcnemp\ are hardly affected by this
choice.

Adopting also the DM91 mass-ratio distribution in Model~A3 gives a
further increase in the CEMP and NEMP fractions, by factors of 1.5 and
2 compared to Model~A1, as this distribution favours the initial mass
ratios with which most CEMP stars and all NEMP stars are born in model
set~A (see Fig.~\ref{fig:map-q-m1}c).  A similar increase in the
number of NEMP stars is seen in model sets~G and H.  On the other
hand, in Model~G3 the number of CEMP stars is somewhat smaller than in
Model~G1 because most CEMP stars now come from binaries with primary
masses 0.85--1.2\,\Msun\ and $q>0.7$, which is disfavoured by the DM91
mass-ratio distribution (Fig.~\ref{fig:map-q-m1}c). The same applies
to Model~H3. In Models~G3 and H3 \fnemp\ and \fcnemp\ increase by a
factor of 2 compared to the default distribution. While less likely to
be correct, these models cannot be excluded with more than 90\%
confidence (Table~\ref{tab:NEMP-fractions}).

Adoption of the \citet{1979_Miller+Scalo} IMF (Models~A4, G4 and H4)
gives somewhat higher CEMP and NEMP fractions than the
\citet{1993_Kroupa} IMF, but the effect is modest: less than factors
of 1.4 and 1.8 for the CEMP and NEMP stars, respectively.  We conclude
that uncertainties in the distributions of the solar-neighbourhood
binary populations affect the CEMP and NEMP fraction and their ratio
by at most a factor of 2.

\subsection{Modified forms of the initial mass function}

Our results for distributions 5 and 6 show the effect of a modified
initial mass function. Model~A5 assumes the IMF suggested by
\cite{2005_Lucatello_IMF} as being required to reproduce the large
CEMP fraction. The fraction of CEMP stars increases by a factor of 2
compared to the KTG93 IMF (Model~A1), as this IMF favours
intermediate-mass stars over low-mass stars
(Fig.~\ref{fig:map-q-m1}b). However, the CEMP fraction still falls
short of the observed value. Our CEMP fraction is smaller than
\citet{2005_Lucatello_IMF} predict because our ranges of initial
primary mass and orbital period that produce CEMP stars are more
restrictive.
With enhanced third dredge-up the increase in the number of CEMP stars
is only a modest factor of 1.3 (Model~G5) to 1.4 (Model~H5).
Models~A5--H5 also show an increased (C)NEMP fraction, by a factor of
4 over models~A1--H1, respectively, which is the direct result of a
larger weight of intermediate-mass stars in this IMF.  Models~G5 and
H5 have $\fnemp\sim$ 0.07--0.08 and $\fcnemp\sim$ 0.03--0.05, which is
still (marginally) compatible with the observed limits and cannot be
excluded with high confidence.  We note that model~H5, which combines
efficient dredge-up, inhibited thermohaline mixing and a modest change
to the IMF, is the only model that comfortably reproduces a CEMP
fraction of about 20\% while remaining consistent with the small
observed number of NEMP stars at $\mathrm{[Fe/H]} > -2.8$.

The effects noted above are much more extreme when we assume the IMF
suggested by \cite{2007_Komiya}, which has a median mass of
$10\,M_\odot$. This IMF gives rise to a substantial CEMP fraction,
even with default input physics (Model~A6), although it remains
smaller than 20\% even in Model~G6: the effect of enhanced dredge-up
is only modest in combination with this IMF.  However, the main effect
of such a top-heavy IMF is an enormous increase in the number of NEMP
stars. One would expect about 20\% of metal-poor stars to be NEMP and
more than 5\% to be CNEMP according to Models~A6 and G6, and these
numbers are even higher (36\% NEMP and 22\% CNEMP) in Model~H6. In
these models \fnemp\ and \fcnemp\ are strongly incompatible with the
observed limits, at least for $\mathrm{[Fe/H]} > -2.8$.

On the other hand, given the much larger apparent \fnemp\ among stars
with $\mathrm{[Fe/H]} < -2.8$ one might more plausibly consider a
top-heavy IMF at these low metallicities. Although we cannot really
compare our models to this low-metallicity sample, we might assume
that the results in Table~\ref{tab:results} are not sensitive to the
metallicity and also apply to $\mathrm{[Fe/H]} < -2.8$. If this were
the case, then using the analysis of Sect.~\ref{sec:NEMP-stats} we
find the probability of finding at most 7 NEMP stars among the sample
of 26 CEMP and NEMP stars to be only $3\times10^{-6}$,
$4\times10^{-5}$ and $2\times10^{-4}$ for Models~A6, G6 and H6,
respectively.
Considering \fcnemp, the probability of finding 5 CNEMP stars among
the 24 CEMP stars would be 3.5\%, 7\% and 0.16\% for Models~A6, G6 and
H6, respectively. This crude analysis suggests that the data at
$\mathrm{[Fe/H]} < -2.8$ do not support a \citet{2007_Komiya}
top-heavy IMF, unless many non-C-rich NEMP stars have remained
undiscovered.

We investigate the constraints on the IMF further in
Figs.~\ref{fig:IMF-modelA}--\ref{fig:IMF-modelH}, where we convolve
our model calculations for Models~A, G and H, respectively, with a
log-normal IMF of the form (\ref{eq:IMFlognormal}) using the median
mass and width of the distribution as free parameters. The top panels
show contours of the CEMP fraction while the bottom panels show
contours of \fnemp, both as a function of \Mmed\ and \DlogM. The
contours in the lower panels correspond to the 90\%, 99\% and 99.9\%
confidence limits on \fnemp\ and the 99\% limit on \fcnemp\ (dashed
line) for $-2.8 < \mathrm{[Fe/H]} < -1.8$, see
Table~\ref{tab:NEMP-fractions}. This is analogous to the analysis of
\citet{2007_Komiya}, the main difference being that we use our
detailed population synthesis models as input and that we also
consider the NEMP/CEMP fraction \fnemp\ as a constraint.

Fig.~\ref{fig:IMF-modelA} demonstrates that Model~A can only reproduce
the observed CEMP fraction for $\Mmed > 1\,\Msun$ and small values of
\DlogM, i.e.\ only for very narrow, top-heavy forms of the IMF. On the
other hand, the upper limit on \fnemp\ only allows small \Mmed\ and
again favours small values of \DlogM. In order to reproduce both with
our default model, we would require a highly fine-tuned IMF with
$\Mmed \approx 2\,\Msun$ and $\DlogM \la 0.2$, i.e.\ one which only
produces intermediate-mass stars in a small mass range.

These stringent constraints are relieved with Model~G, which produces
many more CEMP stars but essentially the same number of NEMP stars,
and thus a lower \fnemp. In this case a wider range of IMFs with
$\Mmed < 2\,\Msun$ is able to reproduce \fnemp, although a large CEMP
fraction ($>15$\%) still requires a rather top-heavy IMF. In this
case, however, CEMP fractions $>10$\% are also produced by a variety
of broad, bottom-heavy IMFs. There is a band of values of \Mmed\ and
\DlogM, which includes the Miller \& Scalo and Lucatello IMFs, that
allows CEMP fractions $>10$\% with acceptable \fnemp\ values.
In Model~H both the CEMP and NEMP fractions increase with respect to
Model~G, while \fnemp\ remains almost unchanged. This model allows a
range of bottom-heavy IMFs ($\Mmed < 2\,\Msun$) that reproduce both
\fnemp\ and a CEMP fraction of at least 15\%.
If we consider only CNEMP stars (dashed lines in
Figs.~\ref{fig:IMF-modelA}--\ref{fig:IMF-modelH}) the ranges of
allowed IMF shapes do not change substantially, except that for
model~G a range of broad IMFs with $\Mmed > 2\,\Msun$ cannot be firmly
excluded.

We conclude that only bottom-heavy IMFs are potentially able to
reproduce both the observed CEMP fraction and the NEMP/CEMP fraction
among stars with $-2.8 \leq \mathrm{[Fe/H]} \leq -1.8$, provided that
efficient dredge-up of carbon occurs in low-mass AGB stars. On the
other hand, top-heavy IMFs (with $\Mmed > 2\,\Msun$) are unable to
produce a low enough \fnemp, if indeed HBB occurs in metal-poor AGB
stars with $M > 2.7\,\Msun$ as detailed models indicate.

\section{Discussion}
\label{sec:Discussion}

In the following subsections we discuss how various uncertainties in
our models and assumptions could affect our results.

\subsection{The lower mass limit for hot bottom burning}

The likelihood of NEMP star formation depends on which masses of star
undergo hot bottom burning. The higher the minimum initial mass for
the onset of HBB, the fewer NEMP stars will form.  Our results are
based on an adopted lower mass limit for HBB, $M_\mathrm{HBB}$, of
2.7\,\Msun\ in accordance with the results of
\citet{2007_Karakas_yields}. On the other hand, \citet{2011_Suda}
argue that if $M_\mathrm{HBB}$ is larger than 5\,\Msun\ then the
observed NEMP/CEMP ratio for $\mathrm{[Fe/H]} < -2.5$ is compatible
with the top-heavy IMF of \citet{2007_Komiya}.

The lower mass limit for HBB is thus an important quantity.
Unfortunately, there is some uncertainty in this quantity, with
different stellar evolution codes giving different values. At
$Z=10^{-4}$, \citet{2004_Herwig} finds $M_\mathrm{HBB} \approx
3.5\,\Msun$ for his models which include convective overshooting. A
similar value is obtained by \citet{2008_Stancliffe}, though they find
some CN cycling occurring in a 3\,\Msun\ model.
\citet{2010_Karakas_yields} finds mild HBB to occur in a 3\,\Msun\
model at this metallicity. At a higher metallicity of $Z=10^{-3}$,
\citet{2009_Weiss} find hot bottom burning does not set in until the
initial mass is $\geq 5\,\Msun$.

Several groups have made grids of AGB models over a range of masses
and metallicities. It is instructive to compare their results, as this
gives some indication of the uncertainties in the determination of
$M_\mathrm{HBB}$. \citet{2008_Campbell} find that HBB occurs for
masses equal to 2\,\Msun\ and above for all the metallicities they
study: from $\mathrm{[Fe/H]} = -3$ right down to the metal-free
population.  \citet{2009_Lau} find $M_\mathrm{HBB}\geq3$\,\Msun\ for
models without overshooting for metallicities from
$\mathrm{[Fe/H]}=-2.3$ down to $\mathrm{[Fe/H]}=-6.3$, the lowest
metallicity in their grid. When they include convective overshooting
during the pre-AGB evolution $M_\mathrm{HBB}$ drops to 2\,\Msun. In
stark contrast to this is the work of \citet{2010_Suda} who do not
find HBB in {\it any} of their grid of models, which spans
metallicities from $\mathrm{[Fe/H]} = -2$ down to $\mathrm{[Fe/H]} =
-5$ (plus $Z=0$) and masses from 0.8\,\Msun\ to 9\,\Msun.

The origin of these differences is not easy to determine. Certainly
one contributing factor is that different groups use different mixing
lengths. This will change the efficiency of convection and hence the
mass at which HBB sets in. It seems likely that the efficiency of
third dredge-up (one of the perennial uncertainties of AGB modelling)
also affects the occurrence of HBB, with more efficient TDU likely to
cause the onset of HBB at lower masses.  Another possibility is the
treatment of molecular opacities.  \citet{2007_Marigo} found that when
a star reaches $\mathrm{C/O}>1$ then hot bottom burning can be
quenched in some cases. This would shift $M_\mathrm{HBB}$ to higher
masses, although \citet{2007_Marigo} expects the effect to be small
for $Z < 10^{-3}$.

\subsection{Non-canonical evolution}

The preceding discussion has focused solely on canonical models of
stellar evolution, that is to say, models that only transport material
by convection. But it has long been thought that some extra mixing
process may be at work in AGB stars, circulating material from the
base of the convective envelope to the hydrogen burning shell. Whether
this is necessary or not is an open question at higher metallicities
(see \citealt{2010_Karakas} and \citealt{2010_Busso} for both sides of
the argument), but at low metallicity it seems likely that some extra
mixing process is at work.
\citep{2009_Stancliffe_mixing,2010_Masseron,2011_Lucatello}.  The
physical cause of this extra mixing is unknown.
\citet{2010_Stancliffe} has shown that thermohaline mixing (as
calibrated to match the abundance changes on the RGB) is unable to
affect nitrogen production in low-metallicity AGB stars.

The effect that extra mixing would have on nitrogen production is
difficult to assess while the nature of the mechanism remains unknown.
From observations of CEMP stars it seems that only a weak circulation
is required, as these stars still show carbon enhancements in excess
of nitrogen enhancements, and few CEMP stars show $^{12}$C/$^{13}$C
ratios as low as the CNO cycle equilibrium value. So while extra
mixing appears to be needed, it seems unlikely that it will
significantly change the minimum mass for HBB.

At low metallicity, AGB models predict mixing episodes not seen at
higher metallicity. The intershell convection zone is able to
penetrate into the H-rich envelope, sucking protons down to high
temperatures and leading to a large release of energy from hydrogen
burning. These {\it dual shell flash} events (or proton ingestion
episodes) can lead to the intershell convection zone splitting into
two regions. One is driven by helium burning and the other by hydrogen
burning. In the hydrogen-burning driven convection zone, the CN cycle
takes place and potentially the nitrogen abundance can be elevated
when the convective envelope penetrates into this region. Several
groups have recently attempted to model this phase, including
\citet{2008_Campbell}, \citet{2009_Lau}, \citet{2009_Iwamoto},
\citet{2009_Cristallo} and \citet{2010_Suda}.  However, these models
are all based on 1D mixing length theory which is almost certainly
incorrect in these circumstances. What is really needed are
hydrodynamical simulations of this phase of evolution.  Recently,
\citet{2011_Stancliffe} performed such simulations. They found serious
deficiencies in the way 1D evolution codes model the process of proton
ingestion, noting that the transport of hydrogen is not diffusive in
character, and that the velocities of the flows involved are
significantly in excess of those given by mixing length theory.
Crucially, they also found no evidence of the convective region
splitting into two zones, despite significant energy generation from
hydrogen burning.

While dual shell flash events may enhance the nitrogen abundance, it
seems unlikely that they would produce nitrogen-rich AGB stars, as
this would require a large degree of hydrogen ingestion together with
fairly complete CN cycling. Whether this is the case will have to
remain unknown until more detailed multi-dimensional calculations can
be made. In summary, it seems improbable that non-canonical evolution
will produce nitrogen-rich AGB stars over a greater mass range than
currently predicted.

\subsection{Mass transfer efficiency}

In our models mass transfer is treated as a combination of Roche-lobe
overflow (RLOF) and Bondi-Hoyle wind accretion \citep{1944_Bondi}.
The modelling of both processes has substantial shortcomings when
applied to binaries containing AGB mass donors.
Roche-lobe overflow is relevant in binary systems that are close
enough for the AGB star to fill its Roche lobe, and in most cases
leads to unstable mass transfer and a common envelope. This evolution
channel produces only a small fraction of the CEMP and NEMP stars in
our models.  However, the classical Roche geometry that we have
assumed may not apply to binaries containing luminous AGB stars, where
radiation pressure may strongly modify the sizes and shapes of the
Roche lobes \citep{1972_Schuerman,2009_Dermine}. This could affect the
stability of RLOF and the number of CEMP stars and NEMP stars produced
by the RLOF channel.  Because AGB stars undergoing HBB are
intrinsically brighter than lower-mass AGB stars, the effect may be
largest on the formation of NEMP stars.

The dominant formation mechanism of CEMP and NEMP stars in our models
is wind accretion. By adopting Bondi-Hoyle accretion we make the
implicit assumption that the wind velocity exceeds the orbital
velocity. This assumption breaks down when the mass-losing star is an
AGB star in a relatively close orbit. AGB mass loss is driven by a
combination of stellar pulsation and radiation pressure on dust. The
wind is accelerated in the region where dust condenses, at a distance
of several stellar radii.
Hydrodynamical simulations show that when the companion orbits at a
distance where the AGB wind has not yet been accelerated to its
terminal velocity, the gas flow geometry resembles Roche-lobe overflow
rather than a Bondi-Hoyle flow
\citep{2004_Nagae,2009_deValBorro,2007_Mohamed,2011_Mohamed}. In this
transition regime of `wind Roche-lobe overflow'
\citet{2010_Mohamed_PhD} and \citet{2011_Mohamed} find that the
companion can accrete up to 50\% of the mass lost by an AGB star, a
much higher accretion efficiency than expected from the Bondi-Hoyle
mechanism ($\la 10\%$). \citet{2012_Abate} have applied a simple model
for wind-RLOF in a population synthesis study and find that this
mechanism increases the occurrence of CEMP stars by a factor of up to
1.5 compared to Bondi-Hoyle accretion.

Of particular relevance for this study is the question whether a more
realistic treatment of mass transfer would change the NEMP/CEMP ratio,
i.e.\ whether it would affect the formation of NEMP stars in a
different way than that of CEMP stars. This question is difficult to
answer without dedicated mass transfer calculations for these cases.
Such calculation are difficult to perform because mass loss of
low-metallicity AGB stars is poorly understood and is not constrained
by observations.
If radiation pressure on dust plays a similar role as at solar
metallicity, then one might expect a difference between massive AGB
stars that undergo HBB and lower-mass AGB stars that do not.  Low-mass
AGB stars produce carbon-rich dust, which condenses easily at
relatively high temperature and thus close to the star.  On the other
hand, stars that have undergone HBB eject gas that is nitrogen-rich
and poor in C and O, which may form dust much less efficiently and at
much larger distance to the star, although little is known about dust
formation under such circumstances\footnote{An interesting test case
  is provided by $\eta$~Carinae \citep{2008_Gull}.}.
Massive AGB stars may therefore accelerate their winds at larger
distance than their lower-mass counterparts. In the framework of the
wind-RLOF model, one may thus expect a larger increase in the number
of NEMP stars than in the CEMP number. If this presumption is correct,
the observed NEMP/CEMP ratio would constrain any modifications to the
IMF even more strongly than we find in Sect.~\ref{sec:Results}.

\subsection{The binary fraction}

The results we present are based on an assumed binary fraction of
100\%, i.e.\ we assume that the progenitor population of CEMP and NEMP
stars consists only of binaries.
For metal-poor halo stars very little is known with certainty about
the multiplicity fraction, but the predominance of binaries among the
CEMP-s stars suggests that binaries are as common in the halo as they
are among Population~I stars in the Galactic disk. It is thus
reasonable to base our discussion on the observed multiplicity in
Galactic disk populations, which is much better determined.
Among nearby solar-type stars, \citet{1991_Duquennoy+Mayor} found a
multiplicity fraction of about 57\% for solar-type stars, which was
revised to $46\pm2\%$ in a more recent analysis of a much larger
sample by \citet{2010_Raghavan}. These studies concern stars of
somewhat lower mass than the progenitors of CEMP stars.  Perhaps more
relevant, certainly for NEMP stars, is the binary fraction among
intermediate-mass stars which was studied by \citet{2007_Kouwenhoven}
in the young OB association Sco~OB2. They found a binary fraction of
at least 70\% (3-$\sigma$ lower limit) and probably much closer to
100\%. A similarly high overall binary fraction is found among massive
O-type stars \citep{2009_Mason,2012_Sana}. A comparison of these
results suggests that the binary fraction among Galactic disk stars is
an increasing function of stellar mass, a conclusion also reached by
\citet{2006_Lada} for low-mass stars. If a similar trend exists for
metal-poor halo stars, then the predicted NEMP/CEMP number ratio from
our models would be higher. This would impose an even stronger
constraint on the IMF.

\section{Conclusions}
\label{sec:Conclusions}

Our results show that the occurrence of NEMP stars, in particular the
observed ratio of NEMP to CEMP stars, sets important constraints on
the initial mass function of the early Galactic halo. In the
mass-transfer scenario, the production of nitrogen by hot bottom
burning in intermediate-mass AGB stars implies that any shift in the
IMF towards more massive stars is accompanied by an increase in the
NEMP/CEMP ratio. From the currently known census of 12 candidate NEMP
stars with $\log g \leq 4$ we derive an upper limit to the NEMP/CEMP
ratio of 0.15 (at 99\% confidence). In the metallicity range $-2.8
\leq \mathrm{[Fe/H]} \leq -1.8$ only two candidate NEMP stars are
known, and we find NEMP/CEMP $<0.09$ with 99\% confidence.  Our
detailed AGB models at $\mathrm{[Fe/H]}=-2.3$, in which HBB occurs in
AGB stars with $M > 2.7\,\Msun$, exclude IMFs with median masses of
$\ga 2\,\Msun$ for stars formed in this metallicity range.
On the other hand the observed NEMP/CEMP ratio is quite compatible
with the solar-neighbourhood IMF, provided that low-metallicity AGB
stars in the mass range 0.85--1.2\,\Msun\ are capable of dredging up
carbon and turning a low-mass binary companion into a CEMP star, as we
concluded in Paper~I.

The sample of known NEMP stars mostly have $\mathrm{[Fe/H]}<-2.8$,
which suggests that the NEMP/CEMP ratio is a strong function of
metallicity. At $\mathrm{[Fe/H]}<-2.8$ the observed NEMP/ CEMP ratio
is approximately 0.27 (with a 99\% confidence upper limit of 0.51).
From currently available model sets of AGB stars it is not clear how
the mass limits for HBB behave at such low metallicities, and
therefore it remains unclear exactly what the implications are for the
IMF among the most metal-poor halo stars. Assuming HBB occurs for
similar masses as at $\mathrm{[Fe/H]}=-2.3$, the observed NEMP/CEMP
ratio allows -- and possibly requires -- a shift in the IMF towards
intermediate-mass stars. Nevertheless, a top-heavy IMF such as
suggested by \citet{2007_Komiya} remains firmly excluded, unless HBB
essentially shuts off at the lowest metallicities \citep{2011_Suda}.
A larger census of NEMP stars would be very valuable to put firmer
constraints on possible changes to the IMF at low metallicity.

\begin{acknowledgements}
  We thank the anonymous referee for valuable comments that helped us
  improve the paper.
  RGI was funded by a Marie-Curie fellowship while in Brussels. The
  research leading to these results has received funding from the
  Seventh Framework Programme of the European Community under grant
  agreement 220440. RJS is a Stromlo fellow. During his time at Monash
  University, he was funded through the Australian Research Council
  Discovery Projects scheme, under grant DP0879472.  EG acknowledges
  support by a NWO VENI fellowship in Nijmegen.
\end{acknowledgements}


\bibliographystyle{aa}
\bibliography{nemp-refs}

\Online

\appendix

\section{Comments on individual NEMP stars}
\label{sec:NEMP-individual}

In this section we comment on the reported observations for our
potential NEMP stars. If these stars are truly polluted with material
from hot-bottom burning companions they are expected to be rich in
nitrogen, sodium and magnesium (e.g. \citealt{2009_Stancliffe}).
Further processing of C and N because of extra mixing may occur in the
star if is evolved beyond first dredge up.

Alpha elements such as calcium and titanium are enhanced by about
0.3--0.4\,dex in normal Galactic halo stars. Some lithium enhancement
is also expected because of hot-bottom burning although there are many
other processes through which it may be made (\citealt{2009_Iwamoto},
\citealt{2010_Stancliffe}). Heavy s-process elements such as barium
and lead are not thought to be produced in intermediate-mass AGB stars
so are not expected to be significantly enhanced in NEMP stars
\citep{2012_Lugaro}.

The typical $1$-$\sigma$ errors on abundance measurements
$\mathrm{[X/Fe]}$ are 0.2--0.3\,dex.

\subsection{CS22949-037}

%
%
%
%
%
%

The abundances of carbon and nitrogen in CS22949-037 are reported by
\citet{2001_Norris}, \citet{2002_Depagne}, \citet{2004_Cayrel} and
\citet{2008_Cohen} who all find $\mathrm{[N/Fe]}>1$,
$\mathrm{[C/Fe]}>1$ and $\mathrm{[N/C]}>0.5$. CS22949-037 is thus a
CEMP and NEMP star.  \citet{2006_Spite} find
$^{12}\mathrm{C}/^{13}\mathrm{C}=4$ which implies that the CN cycle is
in equilibrium. These abundances are consistent with an intermediate
mass AGB star undergoing hot bottom burning.  The gravity of this
star, $\log g = 1.5-1.7$, implies that it has passed through first
dredge-up and is at the point where extra mixing begins but has not
yet had time to significantly alter surface abundances of carbon and
nitrogen (see Figs. 2, 6, and 7 in \citealp{2009_Stancliffe_mixing}).

Enhancements of sodium ($\mathrm{[Na/Fe]}=+1.57$,
\citealt{2007A&A...464.1081A}) and magnesium
($\mathrm{[Mg/Fe]}=+1.55$, \citealt{2010A&A...509A..88A}) also suggest
hot bottom burning is active. Barium is deficient ($\mathrm{[Ba/Fe]} =
-0.84$, \citealt{2001_Norris}; $-0.66$, \citealt{2008_Cohen}; $-0.5$,
\citealt{2011A&A...530A.105A}) and strontium is mildly enhanced
($\mathrm{[Sr/Fe]}=+0.18$, \citealt{2008_Cohen}; +0.17,
\citealt{2011A&A...530A.105A}). These abundances fit the canonical
view that most s-process enhancement occurs in low-mass AGB stars.

\subsection{CS22960-053}

This star is enriched in carbon and nitrogen with
$\mathrm{[C/Fe]}=+2.05$ and $\mathrm{[N/Fe]}=+3.06$ according to
\citet{2007_Aoki} but in contrast \citet{2007_Johnson} find
$\mathrm{[C/Fe]}=\mathrm{[N/Fe]}=+1.15$. The discrepancy in nitrogen
abundances may well be due to the respective use of CN and NH as
abundance indicators. We can only conclude that this system may be a
NEMP star. The star is also enhanced in magnesium
($\mathrm{[Mg/Fe]}=0.65$) and barium ($\mathrm{[Ba/Fe]}=0.86$).

\subsection{CS29528-041}

If we assume that the molecular abundance measurements of
\citet{2006_Sivarani}, $\mathrm{[CH/Fe]} = 1.59$,
$\mathrm{[CN/Fe]}=3.07$ and $\mathrm{[NH/Fe]} = 3.00$, are indicative
of elemental abundances then this star satisfies our (C)NEMP criteria.
With $\log g=4.0$ it is unevolved and there is no mixing on the RGB.
Sodium is enhanced, $\mathrm{[Na/Fe]} = 1.20$, but magnesium is not,
$\mathrm{[Mg/Fe]} = 0.40$, especially when compared to other alpha
elements such as calcium and titanium which show similar enhancements.
CS29528-041 is rich in lithium, $\log \epsilon_\mathrm{Li} = 1.71$,
and barium $\mathrm{[Ba/Fe]} = 0.97$.

\subsection{CS30314-067}

CS30314-067 classifies as a NEMP star according to the abundances
determined by \citet{2002_Aoki}, $\mathrm{[C/Fe]}=+0.5$ and
$\mathrm{[N/Fe]}=+1.20$, but \citet{2007_Johnson} instead find
$\mathrm{[C/Fe]}=+0.25$ and $\mathrm{[N/Fe]}=+0.50$. Some of the
discrepancy may result from the different nitrogen abundance
indicators used and the different adopted gravities, although the
\citet{2002_Aoki} study also has greater resolution.


\subsection{CS30322-023}

\citet{2006_Masseron} have suggested that CS30322-023 is an AGB star
because its gravity is low, $\log g=-0.3$, but this is contradicted by
\citet{2007_Aoki} who claim $\log g=1.0$. In either case extra mixing
on the RGB may have occurred.  Both \citet{2006_Masseron} and
\citet{2007_Aoki} claim $\mathrm{[C/Fe]} = +0.6$ and significant
nitrogen enrichment ($\mathrm{[N/Fe]}=+2.81$ and $+2.47$ respectively)
which suggest this is a NEMP star. Its abundances of sodium, magnesium
and heavy elements are also enhanced ($\mathrm{[Ba/Fe]}\approx+0.6$
and $\mathrm{[Pb/Fe]} = +1.49$).

\subsection{HD25329}

HD25329 is an unevolved star ($\log g = 4.6$) with
$\mathrm{[C/Fe]}=0.1$, $\mathrm{[N/Fe]}=+1.0$ and
$^{12}\mathrm{C}/^{13}\mathrm{C}>40$ (\citealt{2000_Fulbright},
\citealt{2000_Gratton} and \citealt{2003_Gratton}). Its only slight
enhancements in sodium and magnesium ($\mathrm{[Na/Fe]}=+0.24$ and
$\mathrm{[Mg/Fe]}=+0.59$; \citealt{2003_Gratton}) suggest it has not
been polluted by a hot bottom burning AGB star. The origin of nitrogen
in this object remains unexplained.

\subsection{HD206983}

The metallicity of HD206983 is rather large, $\mathrm{[Fe/H]}=-0.99$
\citep{2010_Masseron}, suggesting it is a nitrogen-rich CH-star rather
than a NEMP star. This is supported by an enhancement of barium,
$\mathrm{[Ba/Fe]}=+0.92$ \citep{2010_Masseron}.

\subsection{HE0400-2030}

The carbon- and nitrogen-rich star HE0400-2030, with
$\mathrm{[C/Fe]}=+1.14$ and $\mathrm{[N/Fe]}=+2.75$ according to
high-resolution observations ($R=50,\,000$) of \citet{2007_Aoki},
qualifies as a CNEMP star. Lower-resolution observations
($R=20,\,000$) of \citet{2006_Lucatello} find less enhancement,
$\mathrm{[N/Fe]}=+1.0$ and $\mathrm{[C/Fe]}=+0.8$. Sodium
($\mathrm{[Na/Fe]} = +0.71$), magnesium ($\mathrm{[Mg/Fe]}=+0.62$) and
barium ($\mathrm{[Ba/Fe]}=+1.64$) are also enhanced.


If this star accreted from a hot bottom burning companion, then the
accreted material must have been diluted during the main sequence
(e.g. by thermohaline mixing, \citealt{2007_Stancliffe}), as the
sodium and magnesium abundances are lower than predicted by
theoretical models.

\subsection{HE1031-0020}

\citet{2006_Cohen} find $\mathrm{[C/Fe]} = +1.63$ and $\mathrm{[N/Fe]}
= +2.48$ for HE1031-002, i.e. it is a CNEMP star. With $\log g=2.2$ it
is a moderately evolved giant and should have finished first dredge up
without significant further extra mixing.  Magnesium is slightly
enhanced $\mathrm{[Mg/Fe]}=+0.5$ but sodium is not measured. Calcium,
heavy s-process elements and lead are enhanced,
$\mathrm{[Ca/Fe]}=+1.12$, $\mathrm{[hs/ls]}>+1$ and
$\mathrm{[Pb/Fe]}=+2.66$, suggest an odd evolutionary history for this
star which does not necessarily involve a hot bottom burning
companion.

\subsection{HE1337+0012}

The most extensive set of abundance measurements for the NEMP star
HE1337+0012 (also known as G64--12) is provided by \citet{2006_Aoki}
who find $\mathrm{[C/Fe]} = +0.49$ and $\mathrm{[N/Fe]} = +1.42$. The
sodium abundance is low, $\mathrm{[Na/Fe]=-1.1}$, so HE1337+0012
probably did not accrete material from a hot bottom burning companion
despite its NEMP status.

\subsection{HE1410+0213}

This star is in the samples of both \citet{2006_Cohen} and
\citet{2010_Masseron} but the authors disagree on its surface gravity
with $\log g=3.5$ and $2.0$ respectively. This gives incompatible
abundances of $\mathrm{[C/Fe]}=+2.33$ and $\mathrm{[N/Fe]}=+2.94$
according to \citet{2010_Masseron} and $\mathrm{[C/Fe]}=+1.83$ and
$\mathrm{[N/Fe]}=+1.73$ from \citet{2006_Cohen}. Given the similar
resolution of the studies it is unclear which is correct.


\subsection{HE1413-1954}

\citet{2006_Lucatello} find this is an unevolved ($\log g > 4$) CNEMP
star with $\mathrm{[N/Fe]} = +2.5$ and $\mathrm{[C/Fe]} = +1.7$.
\citet{2005_Barklem} measure $\log g=4.59$, $\mathrm{[C/Fe]}=+1.45$
and $\mathrm{[Sr/Fe]}=-0.47$. The lack of further abundance
measurements prevents a reliable determination of the origin of the
nitrogen in this star.

\subsection{HE2150-0825 and HE2253-4217}

HE2150-0825 and HE2253-4217 are NEMP stars according to the carbon and
nitrogen abundances measured by \citet{2006_Lucatello} who find
$\mathrm{[C/Fe]}=+0.3$ and $\mathrm{[N/Fe]}=+1.3$, and
$\mathrm{[C/Fe]}=+1.0$ and $\mathrm{[N/Fe]}=+1.5$ respectively. The
status of HE2253-4217 as a NEMP is marginal with $\mathrm{[N/C]} =
0.5$, while HE2150-0825 has a low total CN abundance,
$\mathrm{[(C+N)/Fe]} = 0.76$.

\subsection{CS29528-028 and SDSS1707+58}

Neither CS29528-028 nor SDSS1707+58 has measured nitrogen, so they are
not classed as NEMP stars, but their sodium and magnesium abundances
are unusual.  CS29528-028 is an unevolved star with $\log g=4.0$.
\citet{2007_Aoki} report that CS29528-028 has $\mathrm{[Na/Fe]}=+2.68$
and $\mathrm{[Mg/Fe]} = +1.69$ despite a normal calcium abundance
($\mathrm{[Ca/Fe]}=+0.46$).  Its barium abundance is
$\mathrm{[Ba/Fe]}=+3.27$.

SDSS1707+58 is similarly enhanced with $\mathrm{[Na/Fe]}=+2.71$ and
$\mathrm{[Mg/Fe]}=+1.13$ as well as $\mathrm{[Ba/Fe]}=+3.40$
\citep{2008_Aoki}.  This star was recently found to be an RR Lyr star
by \citet{2012_Kinman}, who also find a lower metallicity
$\mathrm{[Fe/H]}=-2.92$ and somewhat lower sodium and barium
abundances.

Based on their sodium and magnesium abundances, these stars are quite
possibly polluted by a companion undergoing hot bottom burning,
although their barium enhancements do not support this view.
Unfortunately, until their nitrogen abundance and carbon isotopic
ratios are measured the status of these stars remains unknown.

\end{document}